\documentclass[hidelinks,times,12pt,a4paper]{article}

\usepackage{lmodern}	
\usepackage[utf8]{inputenc}					
\usepackage{graphicx}						
\usepackage{graphics}
\usepackage{amssymb}
\usepackage{amsmath,amsthm,amssymb}
\usepackage{mathrsfs}	
\usepackage[top=2cm,inner=1.5cm,outer=1.5cm,bottom=2cm]{geometry}
\usepackage{natbib}
\usepackage{comment}

\usepackage{subfigure}
\usepackage{rotating}
\usepackage{color}
\usepackage{hyperref}
\usepackage{tikz}
\usepackage{comment}
\usepackage{setspace}

\usepackage{pagenote}

\usepackage{titling}
\makepagenote

\begin{document}


\pagestyle{myheadings}
\setcounter{page}{0}

\author{Jos\'{e} Raimundo Carvalho\thanks{Author for correspondence. Professor in the Graduate Program in Economics, CAEN/UFC and Director of the Colégio de Estudos Avançados, CEA/EIDEIA. Prédio da EIDEIA – Escola Integrada de Desenvolvimento e Inovação Acadêmica, Campus do Pici S/N - CEP 60455-760 - Fortaleza – CE, Brazil. E-mail: josecarv@ufc.br. Telephone: +5585-991232037.} \and Diego de Maria André\thanks{Departament of Economics, Federal University of Rio Grande do Norte (UFRN). Av. Sen. Salgado Filho, 3000, Lagoa Nova, Natal/RN, CEP 59078-970, Brazil. E-mail:diego.andre@ufrn.br. Telephone: +5584-3215-3509} \and Yuri Costa\thanks{CIA HERING, S\~{a}o Paulo, Brazil}}

\title{Individual Updating of Subjective Probability of Homicide Victimization: a ``Natural Experiment'' on Risk Communication}

\date{}
\maketitle

\noindent \textbf{Abstract}

\noindent We investigate the dynamics of the update of subjective homicide victimization risk after an informational shock by developing two econometric models able to accommodate both optimal decisions of changing \textit{prior} expectations which enable us to rationalize \textit{skeptical} Bayesian agents with their disregard to new information. We apply our models to a unique household data ($N = 4,030$) that consists of socioeconomic and victimization expectation variables in Brazil, coupled with an informational ``natural experiment'' brought by the sample design methodology, which randomized interviewers to interviewees. The higher \textit{priors} about their own subjective homicide victimization risk are set, the more likely individuals are to change their initial perceptions. In case of an update, we find that elders and females are more reluctant to change \textit{priors} and choose the new response level. In addition, even though the respondents' level of education is not significant, the interviewers' level of education has a key role in changing and updating decisions. The results show that our econometric approach fits reasonable well the available empirical evidence, stressing the salient role heterogeneity represented by individual characteristics of interviewees and interviewers have on belief updating and lack of it, say, skepticism. Furthermore, we can rationalize \textit{skeptics} through an informational quality/credibility argument.



\newpage

\section{INTRODUCTION}\label{Section: INTRODUCTION}

The last four decades of economic research are characterized by major advances in applied microeconomics. Issues previously restricted to other social sciences have been included into economists' research agenda due to an increasing interest in interdisciplinary topics. This is the context in which the economics of crime emerged with the seminal paper of \citet{Becker1968}. In that paper, he models the engagement in illegal activities as the result of rational decisions following a cost/benefit analysis. Based on this idea, discussions on deterrence and punishment strategies were raised assisting the design of public policies for crime control. 

Beside criminals, delinquency also involves a victim. In this sense, restricting the analysis of crime on the criminals' perspective is not exhaustive: the victim's side should also be put under scrutiny. Although it is important to stress that becoming a victim is a rare event specifically focused on certain socioeconomic groups, the fear of being victimized, i.e. the subjective expectation of becoming a victim or the ``fear of crime'', plays a significant role in welfare. 

Indeed, the issue of ``fear of crime\pagenote{For our purposes, we adopted the notion advocated by \citet{Warr2000} who defines the ``fear of crime'' as an emotion, a feeling of alarm or fear caused by an awareness or expectation of danger. Therefore, fear of crime affects people negatively, many times more than crime itself.}'' (sometimes denominated ``sensation of (un)safety'') and the correlate concept of risk (future probability) of victimization occupy an important space not only in scientists' agendas but also of public safety managers (\citet{Warr2000} and  \citet{Ferraro1995}). As a consequence, policies and actions to manage crime and the fear of crime must be handled as distinct social problems (\citet{Ferraro1995} and \citet{Vanderveen2006}), although intrinsically related. Thus, given the importance of fear and the weak probabilistic justification for such a severe concern, understanding the factors underlying crime expectations and how information affect those risk perceptions, or ``fear of crime'', is a gap we want to fill in.

In summary, this paper, much in the spirit of the literature initiated by \citet{Viscusi1979} in a labor market context, assumes that citizens are rational agents with imperfect information on the actual probability of becoming a crime victim. Nevertheless, they might face information about crime, not necessarily credible, in such a manner compatible with a learning process. In light of this understanding, we want to study subjective expectations in the context of victimization, investigating the role of informational shocks on posterior risk perceptions, say, ``fear of crime''. \textbf{More specifically, we aim to look into the factors working on the update of individuals' ``fear of crime'' after they receive a shock of information regarding official crime rates.}

For this task, we seek to develop an estimable model able to accommodate both optimal decisions on changing or keeping the same initial perception with respect to subjective homicide victimization risk. This enables us to rationalize the lack of use of new information, contributing to the literature by allowing the existence of non-updaters in a bayesian updating context. Furthermore, we search for observable characteristics to explain this two-step decision, i.e. i) Changing or keeping initial perception; ii) Setting an update (if any).

The bayesian approach proposes a link between current and past beliefs based on further evidence. Hence, situations involving initial perceptions followed by informational shocks are ideally suited for such analysis. However, since this approach fundamentally deals with subjective probabilistic beliefs, two relevant concerns arise: i) Are individuals able to think in terms of probabilities? ii) Are those responses reliable? Despite these concerns, there is strong evidence supporting the bayesian approach in situations where respondents face relevant life aspects, such as becoming victimized by homicide. 

This is the case, for instance, appearing in \citet{Viscusi1984}. They investigate if chemical industry workers learn about the risks they face on the job and how this understanding affects their reservation wages. They conclude that the manner workers update their probabilistic beliefs is in compatibility with a bayesian procedure. This main result is also found in \citet{Viscusi1985}. In addition, an application of this technique in health economics can be found in \citet{Smith1988}: they analyze agents' ability to process information about the health risks of being exposed to a harmful substance. The main message brought by this paper is that information neither recent nor relevant to individuals has insignificant effects on risk perceptions. On the other hand, information seen as more updated and relevant appears to be positively correlated with the posterior probability response.

The dataset used in our paper also presents strong evidence to back up that last statement. Our data derives from a household survey conducted in the city of Fortaleza, Brazil, between October 01, 2011 and January 19, 2012. Twenty interviewers applied a questionnaire to 4,030 citizens of Fortaleza dealing with subjective expectations about being a victim of homicide on the forthcoming twelve months, among other issues. Apart from its size and detailed information on both interviewees and interviewers, perhaps the most distinguished feature coming out of our sample is that the vast majority of individuals (95\%) kept their priors about the risk of being victimized. This is so in spite of the fact that their perceptions were far from official rates and that they received information about the true value. We will refer to this behavior as \textit{skepticism} and to these respondents as \textit{skeptical agents}. 

Instead of questioning the rationality of individuals who decided not to update expectations in a learning context (\textit{skeptical agents}), we decided to approach our empirical evidence slightly different: we might be facing skeptical agents whose perceived new informational content lacks the minimum level of relevance and/or trust to (rationally) prompt any change in risk perceptions. Our results support such understanding. For instance, in every model we estimated, interviewers' level of education, a central variable concerning informational quality and credibility, has positive estimated coefficients regarding the changing of subjective expectations decision. The same is implied by a matching variable in gender, whose positive coefficients may tell us that homophily can either facilitate understanding in communication or signal trustworthy information -- or both.

Our paper is comprised of five more sections. Section \ref{Section: LITERATURE} presents a brief discussion on subjective expectations and Bayesian update. Then, it analyses the contributions of \citet{Viscusi1984} seminal paper and its influence on \citet{Smith1988} with great detail. Finishing the section, a recent application of subjective expectations in \citet{Delavande2008} is presented. Section \ref{Section: THE DATA} presents the data, emphasizing its advantages in comparison with previous papers and our major challenge: to deal with \textit{Skeptical Agents}. Also, it brings an initial exploratory analysis, as well as a brief discussion on selection bias. Econometric models are presented in Section \ref{Section: Econometric Models for Bayesian Updating}. An extension of \citet{Viscusi1984}'s and \citet{Smith1988}'s models with a multiple linear regression is found in subsection \ref{Subsection: Multiple Linear Regression and Victimization Expectation Update}. Subsection \ref{Subsection: Bayesian Updating with Skepticism} develops two models to overcome the theoretical difficulties raised by our data and Section \ref{Section: RESULTS} shows all estimation results. Finally, Section \ref{Section: CONCLUSIONS} concludes, summarizing major messages and proposing further directions of research.

\section{LITERATURE REVIEW} \label{Section: LITERATURE}

\subsection{Subjective Expectations and Bayesian Update} \label{Subsection: Bayes's Theorem and Subjective Expectations}
Given the connection between initial perceptions and further evidence, the bayesian appeal for situations where agents are exposed to new informational bits is straightforward. Nevertheless, a previous step for its use is of great resistance in economics and econometrics: traditionally, scholars have been denying the necessary credibility\pagenote{See \citet{Machlup1946} and \citet{Hart1960}.} to information not objectively observed . Advocates of such understanding base their approach on \citet{Samuelson1938}'s point of view that the only information needed and usable is that of behavior\pagenote{An example of this acquaintance in Econometric Theory is found in \citet{McFadden1973}.}. Proceeding this way, \citet{Samuelson1938} suggests that it is possible to recover initial preferences and the risk of making wrong assumptions about the introspection process is avoided. Also, it prevents any inconsistency raised by possible differences between what is said and done by the decision maker. 

In this sense, whenever it is necessary to comment on the expectations formation process, the rational expectation assumption is usually taken for granted, or, at least, bounded rationality. However, as pointed out by \citet{Pesaran1987}, rational expectations has not proven to be sufficiently convincing to analysis out of macroeconomic long-run even in its weaker version. By the same token, \citet{Manski2004}, when looking at returns in a schooling context, argues that students face the same inferential problems that labor econometricians do. Therefore, it is implausible to assume that students share the same objective and correct distribution function.

The problem is that by relaxing the assumption of a unique objective probability distribution, perfectly known and identically processed by all individuals, the standard analysis cannot take place. This is so because any specific choice behavior may arise as a consequence of many different preferences and probabilities combinations: a classical identification problem. Proposing a different perspective, \citet{Manski2004} presents a detailed defense in favor of using expectations self-reported in probabilistic terms, the so called \textit{subjective probabilities}\pagenote{A first attempt to use subjective probabilities in economics is due to \citet{Juster1966}, though.}. This data type could then be used to validate or relax hypothesis used in traditional models, i.e. those based on objective probabilities only.

In light of such perspective, several papers have successfully applied the concept of subjective probabilities in many situations where expectations are crucial in their analysis. For example, Manski have co-authored studies in income expectations (\citet{Dominitz1997}), social security and retirement (\citet{Dominitz2002}) and consumer confidence (\citet{Dominitz2004}), among others. Also, an application in survival expectations is presented in \citet{Hurd1995}. Once we have brought the concept of subjective probabilities up to the surface, criticisms about the reliability of that data type started to appear. However, before we arrived at that junction we have to go over the important early contributions of W. Kip Viscusi on applying bayesian update approaches to model updating risk perceptions. 

\subsection{Early Contributions}

\citet{Viscusi1979}, interested in the relation between risks for health and physical integrity at work and labor market outcomes, finds that workers' perceived risk has a positive correlation with industry specific risks. Also, he shows that industries whose employees have higher risk perceptions present higher wages as well, supporting the theory of compensating differentials . Therefore, understanding how risk perception evolves and how it is processed emerges as a further step on the study of labor markets dynamics. 

\citet{Viscusi1984} investigate whether employees revise their risk perceptions and, if so, how the reservation wage revision under a new informational set would be. The authors applied a questionnaire to 335 employees of three major American chemical companies. Workers were asked to mark their risk perception in a linear scale ranging from very safe to dangerous, with the US private sector accident and diseases rate presented as a reference point. Finally, answers were converted into probabilistic terms giving rise to the variable \textit{RISK}. 

The first message of \citet{Viscusi1984} is that the data is consistent with a model in which the choice of employment when facing risks follows a learning process. Reservation wages grow as perceived risks increase, so that there is a risk premium keeping workers at their jobs until such compensation is not enough. From that point on, they decide to leave the job or, at least, not to accept the offer if this choice was put back. 
			
The major contribution of this paper, however, is to investigate the informational shock effect on workers' perceptions, which is exactly what we want to do in a victimization context. The authors present each worker a label containing information regarding a chemical substance that would replace the old products used in everyday work activities. Then, workers were asked to repeat their responses under this new information set. Regression results show that workers \textit{posteriors} have been influenced by both \textit{priors} and label contents, creating a reviewing process consistent with a bayesian approach. This is the starting point for our initial estimable model presented in subsection (\ref{Subsection: Multiple Linear Regression and Victimization Expectation Update}).

Motivated by \citet{Viscusi1984}, \citet{Smith1988} investigate how a sample of households in Maine, United States, responded to information regarding risks associated with radon 
concentration in their homes and water supplies. Whereas radon problems in the area had been published widely since the 1970's, the authors seek to address the question more specifically. Besides distributing pamphlets with information about health risks caused by this chemical substance, they informed each household under study the radon concentration in its residence.

The analysis is based only on 117 of 230 observations, those households whose information provided is complete. Although questions may be raised about a possible selection bias, the authors argue that much of the sample loss is due to incomplete information on socioeconomic and behavioral variables not related to one's ability in providing a valid response. The results bring up five central conclusions validating the use of available socioeconomic and environmental information in updating expectations:

\begin{enumerate}
\item [i)] The level of exposure to the radon information provided has a positive impact on the \textit{posterior} magnitude;

\item [ii)] Mitigating actions are negatively related to risk perceptions;

\item [iii)] Cancer victims are very likely to increase \textit{posterior} perception and the weight associated with the radon concentration;

\item [iv)] Life expectancy and age exhibit statistically insignificant results;

\item [v)] Individuals perceive new information with a third of their inital expectations;
\end{enumerate}
		
\noindent Somehow extending the discussion presented in \citet{Viscusi1984} to include other information than \textit{priors}, \citet{Smith1988} provides an important insight for our study. Since the extent on which information is perceived determines its use when updating expectations, we should be cautious with these aspects in our analysis. Also, it is important to note that, despite the empirical success of these articles, we cannot overlook studies observing some nonconformity  between experimental or empirical results and its underlying theory. As pointed out by \citet{Tversky1974}, the literature generally recognizes several specific decision rules (\textit{heuristics}) used unconsciously and not necessarily in an optimal way to solve complex problems.

As it is the case for homicide victimization, dealing with rare events is an even greater challenge for most individuals. As an illustration of this, \citet{Kunreuther1978} shows that risk-averse consumers decided not to purchase insurance against floods and earthquakes even though they had been largely subsidized to do so. However, despite the difficulty in modeling choice behavior in this context, papers such as \citet{Lichtenstein1978} indicates a pattern: regarding risk perceptions, agents tend to overestimate low probabilities and underestimate those of high magnitude. 

Anticipating part of the data discussion found in Section \ref{Section: THE DATA}, our empirical evidence is an example of this pattern. There is a huge gap between respondents' initial perceptions and official values. The sample \textit{prior} mean is strikingly 943 times greater than the official rate for homicide. \citet{Viscusi1985} shows that this result is exactly what one might expect from a bayesian learning process as a final argument for using a bayesian update model in this context. 

\subsection{Recent Developments}
	
More recent papers on subjective expectations focus primarily on the development of probabilistic elicitation mechanisms. More specifically, they aim to assess beliefs in terms of probabilities and to take the extensive evidence about heuristics on the updating process into account. 
Naturally, issues concerning expectation revisions add even more challenges than collecting subjective probabilities. In this context, researchers must be aware of how that reviewing process takes place. 

In an experimental setting, \citet{El-Gamal1995} find evidence on the use of certain heuristics -- \textit{conservatism} and \textit{representativeness} --  in addition to the Bayes's rule when updating subjective probabilities. In general, results have shown so far that this rule is the most likely approach used by agents, but researchers should be alert to the fact that it is not the only one. This message is also found in \citet{Dominitz2011}. Looking at expectation revisions on equity returns, their paper evidences extensive heterogeneity on the updating process.

In light of these results, \citet{Delavande2008} develops the \textit{ERS} (\textit{Equivalent Random Sample}), a metric for subjective expectation updates, in order to accommodate conflicting results about the specific revision process used by individuals. This metric is developed as follows: Let $P_{i}$ be the true probability that an individual $i$ experience the occurrence of $B$, a binary event. Assume that $i$ is unaware of the value of $P_{i}$, so that he has only a subjective expectation about it. Furthermore, let $f_{i, 1} \in \Gamma$ be the distribution function of this initial subjective expectation, where $\Gamma$ is the set of all distribution functions defined on the interval $[0,1]$.

The \textit{ERS} was designed to access posterior subjective probability distributions. Nevertheless, the idea behind it can also be used to discuss informational perceptions and/or credibility. We believe the decision and magnitude of an update is critically dependent on information quality and a no-changing decision perhaps is the best evidence to explore this idea.

\section{DATA} \label{Section: THE DATA}

Before starting the description of our sample, we want to emphasize some key aspects that make us believe in its uniqueness. Firstly, we are dealing with crime risk perception data, a type of empirical evidence hard to find in the literature of risk and information, despite its relevance for social welfare. Furthermore, our sample relates to a developing country, and it is well known that obtaining microdata in such societies challenges researchers of all fields. Focusing on crime issues, according to a recent study conducted by \citet{Drugs2013}, Brazil shows one of the highest homicide rates in the world. 
Therefore, if one wants to study crime perceptions, our data offers rare information of individuals living in a pertinent environment. 

Secondly, as to subjective perceptions, the amount of observations is far greater than what is often presented in some important papers in the field. For example, \citet{Viscusi1984} presents no more than 400 respondents in a seminal paper. Our sample is more than 10 times greater than that, presenting responses of individuals from all of the 116 districts of Fortaleza with an ample range of age, education and income, to name some of the available information. 

In addition, unlike \citet{Delavande2008}, who conducted herself the survey, we have twenty different interviewers. If we want to model information use and updating decisions, having a unique interviewer might introduce selection bias due to the fact that informational source does not change. More importantly, the random variation coming from the sender's side, raising differences in gender, age and educational levels, enables us to investigate aspects of matching with no concerns of self selection, since both participants were not able to choose their matches\pagenote{The interview was held in a randomized household and interviewers were also randomly allocated.}. 

Table (\ref{VariableDefinitions}) presents all variables we are going to use in this study
. Overall, the big picture we are going to discuss in detail is depicted by a middle-age, non-white and female sample, with low income and educational levels. In other words, it is a relevant representation of Fortaleza's residents. 

\begin{table}[h!]
\footnotesize
	\caption{Variable Definitions}
	\label{VariableDefinitions}
	\centering
	\begin{tabular}{lp{7cm}p{3cm}}
		\hline
		Variable & \centering{Description} & Range\\
		\hline
		\textit{Prior} & Initial subjective probability about becoming a victim of homicide on the following 12 months from the interview day onwards. & [0,100] \\
		& & \\
		\textit{Change} & A dummy for respondent decision to change initial perception. & 0: No; 1: Yes.\\
		& & \\
		\textit{Post}\textsuperscript{\textasteriskcentered} & Final subjective probability about becoming a victim of homicide on the following 12 months from the interview day onwards. & [0,100] \\
		& & \\
		\textit{Gender} & A dummy for respondent's Gender.  & 0: Male; \\
		& & 1: Female; \\
		& & \\
		\textit{Age} & Respondent's age in years. & [16,94] \\
		& & \\
		\textit{Police} & A level variable assessing the police work in terms of crime control around the area of respondent residence. & 1: Excellent; 2: Good; 3: Fare; 4: Poor; 5: Terrible. \\
		& & \\
		\textit{Educ\_Int} & A dummy for interviewer's educational level. & 0: High School; 1: More than High School.\\
		& & \\
		\textit{Matching\_Gender}\textsuperscript{\textdagger} & A dummy for the existence of a gender matching between respondent and interviewer. & 0: No; 1: Yes.\\
		\hline
		\multicolumn{3}{l}{{\footnotesize \textsuperscript{\textasteriskcentered} Note that, depending on \textit{Change}, $Post = Prior$ or $Post \neq Prior$.}}\\
		\multicolumn{3}{l}{{\footnotesize \textsuperscript{\textdagger} Note that it does not matter what specific gender matching is observerd.}}\\
		\multicolumn{3}{l}{\footnotesize  Source: Elaborated by the author.}
	\end{tabular}
\end{table}

\noindent Out of 4,030 observations, in which 2,190 are women and 1,840 are men, the average age is slightly more than 39 years old. With respect to marital status, we have 1,928 single individuals, while the portion with a partner sums 2,102. Respondents were also able to identify themselves as White, Black, Mestizo, Asian or Indigenous. Grounded on the standard approach, we classified our sample into White and Non-White, finding only 31\%  as White. In addition, respondents could choose between 8 levels of education and income. Following the Brazilian high income concentration and low educational level, the majority of our sample earns 1 to 2 minimum wage and has an incomplete elementary school degree. 

Getting into our major focus, which concerns the probabilities reported, after a preliminary training in basic probability concepts, participants faced an initial question posed as: \\

\noindent \textit{``Regarding numerical values, what is the chance (or probability) of you being a victim of the following crimes in Fortaleza in the next 12 months: a) Homicide; b) Robbery''}.\\

Despite the controversy surrounding human ability to think probabilistically, and in accordance with \citet{Ferrell1980} and \citet{Koriat1980}, only a bit more than 1/4 of our original sample is not able to give a valid response to this question\pagenote{Those cases are due to participants whose answers are either ``\textit{Didn't Know}'' or ``\textit{Didn't Answer}''.}. We consider this result satisfactory and, due to our large data set, we can eliminate those responses and still keep a large number of observations to work with: 2,885. This first response is referred as \textit{Prior}. 

The interviewer, after receiving this initial probabilistic perception, gives an informational shock as follows: \\

\noindent \textit{``According to the Ministry of Justice, in 2009 the probability of a person in Fortaleza: a) being victim of homicide was 0,037\% (37 homicides for each 100.000 inhabitants) and b) being robbed was 0,96\% (960 thefts for each 100.000 inhabitants)''}. \\

\noindent Interviewees are asked if they want to change their initial responses and, if so, to what value. We call this decision to change as \textit{Change} and the posterior perception as \textit{Post}. 

Table (\ref{Descriptive Analysis - Homicide}) summarizes the sample statistics. We argue that our interaction outcomes rely not only on the respondent's side, but also on the sender's side and on common features between both of them. Since we are trying to model perceived information, we should consider message quality aspects and a first natural choice is the interviewers' level of education. All of our 20 interviewers completed at least high school studies and 80\% of them are undergraduate students or graduates. 

\begin{table}[h!]
\footnotesize
\caption{Descriptive Analysis -- Homicide}
\label{Descriptive Analysis - Homicide}
\centering
\begin{tabular}{lrrrrr}
  \hline
 & Mean & Std.dev & Min & Max & Missing \\
  \hline
Prior & 34.91 & 26.49 & 1.00 & 98.00 & 0 \\
  Change & 0.04 & 0.21 & 0 & 1 & 0 \\
  Post & 33.29 & 26.63 & 0.10 & 98.00 & 0 \\
  Gender & 0.53 & 0.50 & 0 & 1 & 0 \\
  Age & 38.55 & 16.23 & 16 & 94 & 56 \\
  Police & 2.95 & 0.96 & 1 & 5 & 22 \\
  Educ\_Int & 0.83 & 0.37 & 0 & 1 & 0 \\
  Matching\_Gender & 0.77 & 0.42 & 0 & 1 & 0 \\
   \hline
\multicolumn{6}{l}{\footnotesize  Source: Elaborated by the author.}
\end{tabular}
\end{table}

Equally, sources of information noise are concerns to take into account. Note that differences among interviewers and interviewees possibly raise obstacles on information, and it works against an update. In this sense, an important concept to this analysis is \textit{Homophily}. \citet{McPherson2001} defines it as the contact between similar people occurring at a higher rate than among dissimilar people. It is important to keep in mind that, at first, homophily can be related to race, gender, age, religion, social status or any other similar characteristic relating individuals. For example, whenever a group exhibits one of these heterogeneities at a higher rate than it would be found randomly, there is homophily.

The impact of these differences on spontaneous groups’ formation and networks ties is widely documented since the early 1920's. For instance, \citet{Bott1928} noted that school children formed friendships and play groups at higher rates if they were similar on demographic characteristics. The implications of homophily to information diffusion is a research agenda of recent papers such as \citet{Jackson2009} and \citet{Golub2012}. It is worth stressing that one of the main econometrics concern related to matching/group formation is the fact that those matchings are not exogenous. Nevertheless, we are able to bypass this issue here because of the very design of our survey: it imposes double randomization so that the sender - receiver pair is exogenously determined (see, among others, \citet{Graham2011}).

Using this result in our problem, it is possible that the interaction is affected by a particular heterogeneity between interviewers and interviewees. We conjecture that in Latin cultures gender plays a significant role on information credibility/use. Hence, variable \textit{Matching\_Gender} was created to control this issue. In addition, given our interest in victimization risk perception, the way respondents evaluate police forces is a very important piece of information to control for. In a range of five levels, individuals rated the police work around its middle point, which stands for Fare. 

Finally, our subjective probabilities, i.e. \textit{Prior} and \textit{Post}, take almost the full range $[0,100]$ with a huge standard deviation. This initial result piles up to the empirical evidence presented in \citet{Manski2004}, refuting usual concerns about subjective probabilities being reported in values around 50\%. Also, as in \citet{Delavande2011}, we refute the fears that poor, illiterate individuals in developing countries do not understand probability concepts. Equally, the significant difference between respondents's initial perceptions and the official rate could be another source of distrust in such data type. Nevertheless, it finds reverberation in previous papers such as \citet{Dominitz1996} and \citet{BursikJr1999} also attesting the existence of a crime overestimation.

The fact is: although a high discrepancy between subjective victimization probabilities and official rates is not new, our participants were informed about the ``true'' probability and still less than 5\% of them changed initial perceptions\pagenote{ Actually, following \citet{Hoffrage2000}, we presented the same information in numbers (\textit{37 in 100.000}), which brings better results in the participants's use of probabilities.}.Therefore, given that initial perceptions are so overrated, what could explain almost no change in responses?


It is important to keep in mind that we have two different steps after individuals receive an informational shock: i) Whether or not to change initial responses; ii) Given that a change is going to happen, to what value it is going to be set. In a sample of 2885 responders, 2758 of them say no to the first question and 127 behave accordingly to previous studies on bayesian updating processes. Since we have a reasonable updating subsample size here, we can restrict our analysis to this subset of observations and look for any noticeable difference between updaters and the entire sample. Table (\ref{Descriptive Analysis - Updating Subsample}) presents the results.  

\begin{table}[h!]
\footnotesize
	\caption{Descriptive Analysis -- Updating Subsample}
	\label{Descriptive Analysis - Updating Subsample}
	\centering
	\begin{tabular}{lrrrrr}
		\hline
		& Mean & Std.dev & Min & Max & Missing \\
		\hline
		Prior & 39.55 & 28.43 & 1.00 & 90.00 & 0 \\
		Change & 1 & 0 & 1 & 1 & 0 \\
		Post & 2.64 & 2.70 & 0.10 & 15.00 & 0 \\
		Sex & 0.54 & 0.50 & 0 & 1 & 0 \\
		Age & 33.95 & 14.63 & 16 & 78 & 6 \\
		Police & 2.99 & 0.86 & 1 & 5 & 0 \\
		Educ\_Int & 0.91 & 0.29 & 0 & 1 & 0 \\
		Matching\_Sex & 0.93 & 0.26 & 0 & 1 & 0 \\
		\hline
		\multicolumn{6}{l}{{\footnotesize Total of 127 observations.}}\\
		\multicolumn{6}{l}{{\footnotesize Source: Elaborated by the author.}}
	\end{tabular}
\end{table}

Even though we are dealing with a critical question, in which anchoring in expectations is known to be quite frequent, and the amount of information provided might not be ideal, we find that individuals who decide to reevaluate their responses do it in a scathing way. In this case, the subsample set \textit{Post} almost 15 times less than \textit{Prior}. This evidence suggests that our conjecture about the impact of senders' educational level and gender matching into message quality might be on the right path. When we look back to table (\ref{Descriptive Analysis - Homicide}) and compare it with table (\ref{Descriptive Analysis - Updating Subsample}), we find that \textit{Educ\_Int} and \textit{Matching\_Gender} means increased considerably from 83\% and 77\% to $91\%$ and $93\%$, respectively, in our updating subsample. Furthermore, \textit{Prior} and \textit{Age} are other variables that this crude analysis indicates to be aware of. The results show that our updaters are younger individuals with higher initial perceptions. 

In summary, this very simple exercise induces us to believe that a selection bias might be working behind the scenes. Note that the \textit{Post} reported is conditioned to a previous decision and, when we select our sample only for individuals who decide to change initial responses, this subgroup has some different characteristics from the entire sample. Those characteristics might be just what we are looking for.  

In short, the data used here also attests the existence of a crime overestimation found in previous works. The novelty is that our respondents faced an informational shock consisting of the official homicide rate, but 95\% of them  keep the same initial perception. So far, looking at the updating subsample only, we found that these updating interactions were made by more educated interviewers, sharing the same gender with our interviewees. Thus, those two most important results enable us to follow our main line of explanation: informational quality.

In the next section, we start discussing our models. Subsection (\ref{Subsection: Multiple Linear Regression and Victimization Expectation Update}) is the first step of our contribution. We extend \citet{Viscusi1984} and \citet{Smith1988} by proposing a multiple linear regression motivated by victimization expectations. We believe subsection (\ref{Subsection: Bayesian Updating with Skepticism}) is our main achievement. It proposes two alternative models rationalizing the skepticism found in our data and it provides an extension of current models to allow for the existence of skeptical bayesian updaters.

\section{ECONOMETRIC MODELS FOR BAYESIAN UPDATING} \label{Section: Econometric Models for Bayesian Updating}

\subsection{Multiple Linear Regression and Victimization Expectation Update}\label{Subsection: Multiple Linear Regression and Victimization Expectation Update}

Consider a victimization event, where $y=1$ if the individual becomes a crime victim  and $y=0$ on the contrary. Let $Y \sim Bern(\pi)$, \textit{iid}, and assume that the respondent has a probability distribution $Beta(\alpha,\beta$) about the true, but unknown, probability $\pi$ of becoming a homicide victim. 

In this context, we propose the following interaction structure: the initial respondent's perception, $Prior_{i} \equiv \pi_{i}^{0}$, can be modeled as dependent on his observable characteristics, for instance, age, gender, income, education, etc. Formally, we can define a function $f: \mathbb{R}^{q} \rightarrow [0,1]$ such that $\pi_{i}^{0}=f(\tilde{\mathbf{X}}^{r}_{i})$, where $\tilde{\mathbf{X}}^{r}_{i}$ is the $q$ dimensional heterogeneity vector of individual $i$, the \textbf{receiver (interviewee)}. When asked about his initial perception, he assess $P(y=1|\tilde{\mathbf{X}}^{r}_{i})=E(\pi|\tilde{\mathbf{X}}^{r}_{i})=\pi_{i}^{0}$ and establishes $\pi_{i}^{0}=\frac{\alpha_{i}}{\alpha_{i}+\beta_{i}}$ as response. 

After collecting $\pi_{i}^{0}$ by means of the questionnaire, \textbf{sender (interviewer)} $j$  sends an informational bit consisting of the true population value, $\pi$ (homicide prevalence rate in Fortaleza). However, from the respondent's perspective, this information may lack credibility and its content may not be fully taken into account. In fact, let $\pi^{*}_{i}$ be the proper value depicted by this information (from the perspective of the receiver) and define $g:\mathbb{R}^{k} \rightarrow [0,1]$, with $\pi^{*}_{i}=g(\mathbf{X}_{i}^{r},\mathbf{X}_{j}^{s},\mathbf{X}_{i,j}^{m},\pi_{i}^{0})$, where $\mathbf{X}_{i}^{r}$ and $\mathbf{X}_{j}^{s}$ are vectors of observable heterogeneity for receivers\pagenote{The vector determining $\pi_{i}^{0}$ is not necessarily the same for $\pi^{*}_{i}$.} and senders, respectively, and $\mathbf{X}_{i,j}^{m}$ is a vector of matching aspects.

\begin{figure}[h!]\caption{Time Sequence of Sender-Receiver Interaction}
	\begin{tikzpicture}[scale=1.0]
		\draw [->, thick] (-12.0,0) -- (2.0,0)node[below]{\footnotesize{Time}};
	
		\draw [thick] (-12.0,0) -- (-12.0, 0.5) node[above, color=blue]{\small{Begining of interview}}(-12.0, 1.5);
	
		\draw [dashed] (-11.5,0) -- (-11.5, -0.5) node[text width=2.5cm,align=center, below]{\scriptsize{\textit{\textbf{Prior}}\\$E(\pi|\tilde{\mathbf{X}}^{r}_{i})=\pi_{i}^{0}$}}  (-11.0,-2);
	
		\draw [dashed,color=red] (-8.0,0) -- (-8.0, -0.5) node[text width=2.5cm,align=center, below,color=red]{\scriptsize{\textbf{Information revealed}\\$\pi$}}  (-9.5,-2);
	
		\draw [dashed,color=red] (-4.0,0) -- (-4.0, -0.5) node[text width=3.5cm,align=center, below,color=red]{\scriptsize{\textbf{Receiver's perception}\\$g(\mathbf{X}_{i}^{r},\mathbf{X}_{j}^{s},\mathbf{X}_{i,j}^{m},\pi_{i}^{0}|\pi)=\pi^{*}_{i}$}}  (-7.0,-2);
	
		\draw [dashed] (0.0,0) -- (0.0, -0.5) node[text width=2.5cm,align=center, below]{\scriptsize{\textbf{Posterior}\\$\pi_{i}^{p} = E(\pi|\mathcal{I}_{i})$}}  (-4.0,-2);
	\end{tikzpicture}
\label{TimeSequence}	
\end{figure}

\noindent Figure \ref{TimeSequence} sketches the interaction between sender and receiver so far described. So, we have the following important objects already defined:

\begin{enumerate}
\item $\pi_{i}^{0} = f(\tilde{\mathbf{X}}^{r}_{i})$, respondent's prior about his own victimization risk;

\item  $\pi^{*}_{i}=g(\mathbf{X}_{i}^{r},\mathbf{X}_{j}^{s},\mathbf{X}_{i,j}^{m},\pi_{i}^{0}| \pi)$, receiver's perception from the information revealed by sender $j$ when the latter declares that the real victimization risk is $\pi$;

\end{enumerate}

\noindent Abstractly, the information was translated into a sequence of victimization observations, generating an informational content $\mathcal{I}=(0,1,\dots,0,1)$ with $n_{1}$ values equal 1 and $n_{0}$ equal 0, where $N=n_{0}+n_{1}$. In summary, the respondent, whose initial perception was $\pi_{i}^{0}$, uses the perceived information, $\pi^{*}_{i}$, to generate $\mathcal{I}_{i}$, creating a set of new evidence which is the foundation to update his response towards a posterior perception. This new response, $Post \equiv \pi_{i}^{p}$, is given by $P_{i}(y=1|\mathcal{I}_{i})=E(\pi|\mathcal{I}_{i})=\pi_{i}^{p}$. In this structure:

\begin{equation*}
p_{i}(\pi|\mathcal{I}_{i}) \propto p_{i}(\mathcal{I}_{i}|\pi)p_{i}(\pi)
\propto \pi^{n_{1}}(1-\pi)^{n_{0}}\frac{\Gamma(\alpha_{i}+\beta_{i})}{\Gamma(\alpha_{i})\Gamma(\beta_{i})}\pi^{\alpha_{i} -1}(1-\pi)^{\beta_{i}-1} 
\end{equation*}
\indent Then, 
\begin{equation*}
\pi|\mathcal{I}_{i} \sim Beta(\alpha_{i}+n_{1}, \beta_{i}+n_{0}) 
\end{equation*}
\indent And, 
\begin{equation*}
\pi_{i}^{p} = E(\pi|\mathcal{I}_{i}) = \frac{\alpha_{i}+n_{1}}{\alpha_{i}+\beta_{i}+n_{1}+n_{0}} 
\end{equation*}

\noindent For estimation purposes, $\pi_{i}^{p}$ remains undefined, since it is not possible to identify none of the given parameters. One way to dodge this problem is to realize that $\pi_{i}^{*}=\frac{n_{1}}{N}$ and use the fact that $\pi_{i}^{0}=\frac{\alpha_{i}}{\alpha_{i}+\beta{i}}$ to establish:  

\begin{equation*}
\pi_{i}^{p}=\frac{N(\frac{n_{1}}{N})+(\alpha_{}i+\beta_{i})\frac{\alpha_{i}}{(\alpha_{i}+\beta_{i})}}{\alpha_{i}+\beta_{i}+N} 
\end{equation*}

Therefore, we are able to write our structural model as: 

\begin{equation}\label{ModeloEstruturalClassico}
\pi_{i}^{p}= \frac{N}{\alpha_{i}+\beta_{i}+N}\pi^{*}_{i}+\frac{\alpha_{i}+\beta_{i}}{\alpha_{i}+\beta_{i}+N}\pi_{i}^{0} 
\end{equation}

\noindent Now, if we assume that $g:\mathbb{R}^{k} \rightarrow [0,1]$ can be reasonable approximated by a linear form, 
it allows us to extend \citet{Viscusi1984} and \citet{Smith1988} at the same time by taking not only respondent's heterogeneity into account, but also both interviewer and matching information through a coherent conditioning multiple linear regression as follows:: 

\begin{equation}\label{Equation: Final Multiple Linear Regression}
\pi^{p}_{ij}= \lambda+\gamma\pi^{0}_{i}+\mathbf{X}_{ij}\mathbf{\delta}+ \epsilon_{ij}
\end{equation}

\noindent Where $\pi^{p}_{ij}$ is the posterior of receiver $i$, after receiving information from sender $j$, $\mathbf{X}_{ij}$ is an extended vector of observed heterogeneity containing $\mathbf{X}_{i}^{r},\mathbf{X}_{j}^{s},\mathbf{X}_{i}^{m}$, and $\epsilon_{ij}$ the error term.

\subsection{Bayesian Updating with Skepticism}\label{Subsection: Bayesian Updating with Skepticism}

As exposed in Section \ref{Section: THE DATA}, we have a very different sample in the context of bayesian update. Our data presents an excess of individuals --- our skeptical agents --- who do not update and keep their posterior responses with the same value as their initial perceptions, i.e. $Prior_{i}=Post_{i}$. One might point this issue as an important downside of our study. Also, previous papers present data collection procedures specifically designed to assess subjective probabilities and its revision processes\pagenote{\citet{Delavande2008} and \citet{Delavande2011} are good examples of this.} while our dataset comes from a broad household survey.

Clearly, individuals faced a critical question -- homicide expectation -- about which initial perceptions were too far from the truth. In addition, the information provided lasted less than a couple of minutes. However, even under such circumstances, if we proceed the estimation of equation (\ref{Equation: Final Multiple Linear Regression}) considering only those 127 individuals who changed initial responses, we work on a sample almost 40\% greater than what is presented in \citet{Delavande2008}, for example.

By now, we want to extend the discussion on informational content raised by \citet{Viscusi1984} to move from an extreme, where almost every participant revises initial perceptions, to the other, where the contrary occurs. Is that reasonable to keep considering an economy composed by agents who always believe in the information received for every risky decision to be made? Putting in other words, is all information provided about uncertain outcomes informative enough to make almost every individual change their initial perceptions? Our dataset sends a clear message that this is not the case, and restricting attention only to those updaters would lead us, at least, to waste a huge amount of information. Also, these statistics reported might be misleading, due to selection bias. 

In this study, it is important not to neglect individuals who did not respond to new information, since they carry important clues about the updating process. Still, how to accommodate in the same model rational individuals, facing the same information, but behaving so differently? \textbf{At this point, it is clear for us that our venture and major contribution is to propose a framework compatible to a bayesian update allowing the possibility that no update takes place.} In this sense, we seek to develop an econometric model which is able to address the following issues: a) Agents are bayesian when updating their victimization expectations; b) Observable and non-observable heterogeneity must refer not only to the receiver (interviewee) but also to the sender (interviewer), since matching aspects may influence credibility and information usage; c) The presence of non updating individuals must receive an adequate treatment and be rationalized under a bayesian context.

With that in mind, rewrite (\ref{ModeloEstruturalClassico}) as follows: 

\begin{equation}\label{ModeloEstruturalRevisitado}
\pi_{i}^{p}=\big( 1+ \frac{\alpha_{i}+\beta_{i}}{N}\big)^{-1}\pi^{*} + \big(1 + \frac{N}{\alpha_{i}+\beta{i}}\big)^{-1}\pi_{i}^{0} 
\end{equation}	

Note that, from $\pi_{i}^{0}$, we may interpret $\alpha_{i}+\beta_{i}$ as the quantity of samples initially used to form $Prior_{i}$. Following this reasoning, $\eta_{i}=\frac{N}{\alpha_{i}+\beta_{i}}\in [0,\infty)$ may be interpreted as a measure of the informational quality. In this way,	(\ref{ModeloEstruturalRevisitado}) can be rewritten as: 
\begin{equation}\label{ModeloEstruturalAncoragem}
\pi_{i}^{p}=\big( 1+ \frac{1}{\eta_{i}}\big)^{-1}\pi^{*} + \big(1 + \eta_{i}\big)^{-1}\pi_{i}^{0} 
\end{equation}

This structural form will be crucial to skepticism rationalization. We need to explain individuals setting $\pi_{i}^{p}=\pi_{i}^{0}$, and  it is reasonable to assume that the decision to revise perceptions, i.e to make $\pi_{i}^{p} \neq \pi_{i}^{0}$, is directly related to the credibility and quality of the information obtained. As (\ref{ModeloEstruturalAncoragem}) makes clear, when $\eta_{i}$ approaches zero, or the more irrelevant the information is seen by the decision maker, $\pi_{i}^{p}$ approaches $\pi_{i}^{0}$, which, in the limit, translates into the complete non-existence of an update. In the other extreme, when $\eta_{i}$ approaches the infinity, $\pi_{i}^{p}$ takes exactly the value $\pi_{i}^{*}$. 

In summary, individuals setting $Post_{i}=Prior_{i}$ tell us that informational content was either irrelevant or the source lacks credibility, or both, and there was no reason to change initial perceptions. Ignoring these responses introduces a selection bias since establishing  no changing in perceptions is a rational decision based on an intrinsic optimization process, which might be dependent on observable and unobservable characteristics.

In this situation, it is hard to believe that the now restricted error term has a zero conditional mean and, even if the structural model is linear, OLS procedure leads to inconsistent parameter estimates. One possible approach to deal with these cases is to think of an unrestricted and unobservable \textit{latent} variable underlining the true observations through a specific structure. That is exactly what we will develop next.

\subsubsection{Model 1: A Generalized Tobit for Bayesian Updating}\label{Subsection Generalized Tobit}

Initially, let $Y$ be a random variable of interest and $Y^{*}$ its latent counterpart. then, 

\begin{align*}
Y^{*}_{i} &= \mathbf{X}_{i}\delta + u_{i}, & u_{i}|\mathbf{X}_{i} \sim Normal(0,\sigma^{2})\\
Y_{i} &= max(0,Y^{*}_{i}) 
\end{align*}

\noindent where $\mathbf{X}_{i}$ is a  vector of conditioning variables. 

Adapting this Tobit Model for our purposes, we assume the existence of a latent variable $Post^{*}_{i}$ that will govern the censoring mechanism. We conjecture that $Post^{*}_{i}$ captures the respondents' latent \textit{posterior} probability and, as such, according to equation (\ref{Equation: Final Multiple Linear Regression}), we propose it is a function of the \textit{prior} and of a vector of observable variables  containing both interviewees' and interviewers' characteristics, as well variables representing homophily. Hence, consider the following Tobit: 

\begin{subequations}
\begin{align}
Post^{*}_{i} & = \gamma Prior_{i} + \mathbf{X}_{i}\delta + u_{i}, & u_{i}|Prior_{i},\mathbf{X}_{i} \sim Normal(0, \sigma_{u}^{2}) \label{Proposed Tobit.1}\\
    Post_{i}  & =  \min(Prior_{i}, Post^{*}_{i}) \label{Proposed Tobit.2}
\end{align}
\end{subequations}

\noindent But why the minimum? This is so because, in our sample, all initial perceptions are greater than the official rate. Thus, for the sake of rationality, we must have $Post_{i} \leq Prior_{i}$ $\forall i$. Now, note that: 

\begin{equation*}
Post_{i}  = Prior_{i} - \max(0, Prior_{i} - Post^{*}_{i}) 
\end{equation*}

This means that our Tobit is equivalent to the following ``new'' Tobit: 

\begin{subequations} \label{Model: FinalTobit}
\begin{align}
Prior_{i} - Post^{*}_{i} & = (1 - \gamma) Prior_{i} - \mathbf{X}_{i}\delta + v_{i} \label{Estimated Tobit.1}\\
   Prior_{i} - Post_{i}  & =  \max(0, Prior_{i} - Post^{*}_{i}) \label{Estimated Tobit.2} 
\end{align}
\end{subequations}

\noindent where, as usual, we assume that $v_{i}|Prior_{i},\mathbf{X}_{i} \sim Normal(0, \sigma_{v}^{2})$. 

Given our data, the way we defined equation (\ref{Estimated Tobit.1}) would lead us to work with positive values and, in conjunction with equation (\ref{Estimated Tobit.2}), the model just described is a well-known generalization for the original standard Tobit. Note that differently from the traditional approach, the threshold is not constant along observations but vary in a deterministic way, say, $Prior_{i} = h(\mathbf{X}_{i})$. Actually, the form of $h(\mathbf{X}_{i})$ does not need to be specified, since we build our model conditional on $Prior_{i}$. 

Finally, we just need to run a Tobit where the dependent variable is $(Prior_{i} - Post_{i}) \geq 0$ and the vector of regressors is $(Prior_{i}, \mathbf{X}_{i})$. The estimated parameters are $(1 - \widehat{\gamma}, -\widehat{\delta})$. However, we need $(\widehat{\gamma}, \widehat{\delta})$ and interpret these results by means of equation (\ref{Proposed Tobit.1}). Now refer to equations (\ref{Proposed Tobit.1}) and (\ref{Proposed Tobit.2}) and the interpretation is as follows: 

\begin{enumerate}
\item [i)] Interpreting $\widehat{\gamma}$:

The fundamental result to obtain with respect to $\widehat{\gamma}$, attesting the validity of a bayesian update model, is its statistical significance. In other words, this tells us that $Prior$ influences $Post^{*}$ and, thus, $Post$, i.e. we have a bayesian effect in the sense of \citet{Viscusi1984}.

If $\widehat{\gamma} > 1$, then agents are \textbf{SKEPTICAL}.
An incremental change in $Prior_{i}$ induces a higher incremental change in $Post^{*}_{i}$. Hence, since equation (\ref{Proposed Tobit.2}) is the minimun between them, we have $Post_{i} = Prior_{i}$, i.e., the higher the prior, the less likely to update (skepticism).

If $\widehat{\gamma} < 1$, then agents are \textbf{UPDATERS}.
An increase in $Prior_{i}$ induces a less than proportional increase in $Post^{*}_{i}$. Therefore, we are looking at the updaters, and $Post_{i} < Prior_{i}$, i.e., the higher the prior, the more likely to update.

\item [ii)] Interpreting $\widehat{\delta}$:

If $\widehat{\delta} > 0$, then an increase in $\mathbf{X}_{i}$ implies  a higher $Post^{*}_{i}$ and we observe the same effect that $\widehat{\gamma} > 1$. Analogously, if $\widehat{\delta} < 0$, then lower $\mathbf{X}_{i}$ implies lower $Post^{*}_{i}$. Therefore, it is more likely that we observe an update. 

\end{enumerate}

In order to build the original Tobit likelihood function, we need the following expressions first:

\begin{subequations}
\begin{align}
P(Post_{i} = Prior_{i}|Prior_{i},\mathbf{X}_{i}) & = 1 -  \Phi \left(\frac{(\gamma - 1)Prior_{i} + \mathbf{X}_{i}\delta}{\sigma_{u}} \right) \label{Lik1.1}\\
P(Post_{i} =  Post^{*}_{i}|Prior_{i},\mathbf{X}_{i}) & =  \Phi \left(\frac{(\gamma - 1)Prior_{i} + \mathbf{X}_{i}\delta}{\sigma_{u}} \right)\label{Lik1.2}
\end{align}
\end{subequations}

\noindent As well, we need the density $f(Post_{i}|Prior_{i},\mathbf{X}_{i}, Post^{*}_{i} < Prior_{i})$. But this is just:

\begin{equation*}
g(Post^{*}_{i}|Prior_{i},\mathbf{X}_{i}, Update^{*}_{i} > Prior_{i})
\end{equation*}

\begin{equation}
f(Post_{i}|Prior_{i},\mathbf{X}_{i}, Post_{i} > Prior_{i}) = \frac{\phi\left( \frac{(Post_{i} - \gamma Prior_{i} - \mathbf{X}_{i}\delta)}{\sigma_{u}}\right)}{\sigma_{u}\left(1 - \Phi \left( \frac{\left(Post_{i} - \gamma Prior_{i} - \mathbf{X}_{i}\delta\right)}{\sigma_{u}} \right) \right)} \label{Posterior1.1}
\end{equation}

\noindent Defining the indicator function by $\mathbf{1}_{(\cdot)}$, the likelihood contribution of a given $Post_{i}|(Prior_{i},\mathbf{X}_{i})$ observation is:

\begin{equation*}
\left[1 -  \Phi \left(\frac{(\gamma - 1)Prior_{i} + \mathbf{X}_{i}\delta}{\sigma_{u}} \right)\right]^{\mathbf{1}_{(A)}} \times \left[\frac{\phi\left( \frac{(Post_{i} - \gamma Prior_{i} - \mathbf{X}_{i}\delta)}{\sigma_{u}}\right)}{\sigma_{u}\left(1 - \Phi \left( \frac{\left(Post_{i} - \gamma Prior_{i} - \mathbf{X}_{i}\delta\right)}{\sigma_{u}} \right) \right)} \right]^{\mathbf{1}_{(B)}}
\end{equation*}

\noindent where $A: Post_{i} = Prior_{i}$ and $B: Post_{i} > Prior_{i}$.

\subsubsection{Model 2: A Simple Two-Tiered Hurdle Model} \label{Subsection Simple Two-Tiered Hurdle Model}

An alternative modeling comes from the fact that we have a two-step structure. As mentioned before, individuals decide whether or not change their \textit{priors} and then they give a \textit{posterior} perception. Hence, \textit{Discrete Choice Models} such as \textit{Logit} and \textit{Probit} followed by OLS on the updating subsample emerge as a natural approach. Thus, an alternative to the Tobit model is to dispense with the abstraction of a latent variable $Post_{i}^{*}$ and model only the observed empirical evidence directly.  If we define $\Delta^{update}_{i} \equiv -Update = Prior_{i} - Post_{i}$ as the actual update to deal only with non-negative numbers\pagenote{Remember that in our data set updaters have $Prior_{i} \neq Post_{i}$, $\forall$ $i$.}, we have:

\begin{subequations}
\begin{align}
P(\Delta^{update}_{i}  = 0|Prior_{i},\mathbf{X}_{i}) & = 1 - \Phi(\gamma Prior_{i} + \mathbf{X}_{i}\mathbf{\delta}) \label{subeq3.1}\\
G(Post_{i}) |(Prior_{i}, \mathbf{X}_{i}, \Delta^{update}_{i} > 0) & \sim Normal(\tilde{\gamma}Prior_{i} + \mathbf{X}_{i}\tilde{\delta}, \sigma^{2}) \label{subeq3.2}
\end{align}
\end{subequations}

\noindent Equation (\ref{subeq3.1}) dictates the probability that a real update has not occurred, i.e., $\Delta^{update}_{i} = 0$ and equation (\ref{subeq3.2}) is a conditional model (conditioned on $\Delta^{update}_{i} > 0$) for the posterior probability. Equation (\ref{subeq3.2}) just asserts that after a suitable transformation $G(Post_{i})$, usually but not necessarily $G(Post_{i}) = \log(Post_{i})$, we achieve normality. There are some features about the model just described:

\begin{enumerate}
\item [i)] We could have different explanatory variables for the censoring equation (\ref{subeq3.1}) and equation (\ref{subeq3.2}), as well as estimated parameters for the same explanatory can vary between equations;

\item [ii)]  Estimation of $(\gamma, \delta, \tilde{\gamma},\tilde{\delta}, \sigma^{2})$ is quite simple, say: for  equation (\ref{subeq3.1}), run a probit; and for equation (\ref{subeq3.2}) run a simple OLS for the subsample where $\Delta^{update}_{i} > 0$;
\end{enumerate}

\noindent Estimation of partial effects demands knowledge of $G(\cdot)$. However, if we assume log-normality, $G(Post_{i}) = \log(Post_{i})$, nice results emerge:

\begin{subequations}
\begin{align*}
& E(Post_{i}|Prior_{i}, \mathbf{X}_{i}, \Delta^{update}_{i} > 0)  = \exp\left(\tilde{\gamma} Prior_{i} + \mathbf{X}_{i}\tilde{\delta} + \frac{\sigma^{2}}{2}\right) \\
& E(Post_{i}|Prior_{i}, \mathbf{X}_{i})  = \Phi(\gamma Prior_{i} + \mathbf{X}_{i}\delta) \exp\left(\gamma Prior_{i} + \mathbf{X}_{i}\delta + \frac{\sigma^{2}}{2}\right)
\end{align*}
\end{subequations}

\noindent An import drawback of the hurdle model is the impossibility of testing it against the Tobit specification, at least by simple procedures. On the other hand, a Vuong-type test could be applied to those non-nested hypotheses.

\section{RESULTS} \label{Section: RESULTS}

\subsection{Bayesian Update with Skepticism: Model 1}		

For convenience, equations (\ref{Proposed Tobit.1}), (\ref{Proposed Tobit.2}), (\ref{Estimated Tobit.1}) and (\ref{Estimated Tobit.2}) are shown, again, below:

\begin{align}
Post^{*}_{i} & = \gamma Prior_{i} + \mathbf{X}_{i}\delta + u_{i} \tag{\ref{Proposed Tobit.1}}\\
Post_{i}  & =  \min(Prior_{i}, Post^{*}_{i}) \tag{\ref{Proposed Tobit.2}}\\
Prior_{i} - Post^{*}_{i} & = (1 - \gamma) Prior_{i} - \mathbf{X}_{i}\delta + v_{i} \tag{\ref{Estimated Tobit.1}}\\
 Prior_{i} - Post_{i}  & =  \max(0, Prior_{i} - Post^{*}_{i}) \tag{\ref{Estimated Tobit.2}}
\end{align}

Table (\ref{Tobit Model Results}) presents the estimation results. As explained in subsection (\ref{Subsection Generalized Tobit}), when referring to the latent equations (\ref{Proposed Tobit.1}) and (\ref{Proposed Tobit.2}), we see the likelihood of an update. This is the first decision step going on. Keep in mind that, for this part, we are interested in the coefficients of equation (\ref{Proposed Tobit.1}) but, in fact, we estimate equation (\ref{Estimated Tobit.1}). Thus, we need to proceed an adjustment on coefficients interpretation. For the $Prior_{i}$ coefficient, our interest relies on $\gamma$ itself, but we would get an estimative of $(1-\gamma)$; therefore, it is easy to see that we must subtract it from 1. For all of the other coefficients, represented by $\delta$, simply change estimatives' signs and, then, interpretation follows usual reasoning. 

\begin{table}[h!]
\footnotesize
\centering
  \caption{Tobit Model}
  \label{Tobit Model Results}
\begin{tabular}{@{\extracolsep{5pt}}lc}
\\[-1.8ex]\hline
\hline \\[-2.5ex]
 & \multicolumn{1}{c}{\textit{Dependent variable:}} \\
\cline{2-2}
\\[-1.8ex] & (Prior - Post) \\
\hline \\[-1.8ex]
 Prior & 0.502$^{***}$ \\
  & \footnotesize{(0.129)} \\
  \\[-2.5ex]
 Age & $-$0.728$^{***}$ \\
  & \footnotesize{(0.241)} \\
  \\[-2.5ex]
 Sex & $-$18.036$^{**}$ \\
  & \footnotesize{(7.684)} \\
   \\[-2.5ex]
 Matching\_Sex & 63.480$^{***}$ \\
  & \footnotesize{(13.931)} \\
    \\[-2.5ex]
 Educ\_Int & 19.307$^{*}$ \\
  & \footnotesize{(11.459)} \\
   \\[-2.5ex]
 Constant & $-$232.149$^{***}$ \\
  & \footnotesize{(39.056)} \\
\hline \\[-1.8ex]
\footnotesize{Observations} & \footnotesize{2,828} \\
\footnotesize{Log Likelihood} & \footnotesize{$-$1,017.116} \\
\footnotesize{Wald Test} & \footnotesize{41.278$^{***}$ (df = 5)} \\
\hline
\hline \\[-1.8ex]
\multicolumn{2}{l}{\footnotesize{$^{*}$p$<$0.1; $^{**}$p$<$0.05; $^{***}$p$<$0.01}} \\
\multicolumn{2}{l}{\footnotesize{Source: Elaborated by the author.}} \\
\end{tabular}
\end{table}

\noindent The first result we must obtain is the statistical significance of the $Prior$ coefficient in equation (\ref{Proposed Tobit.1}). Table (\ref{Tobit Model Results}) shows that $\hat{\gamma}=(1-0,5021)= 0.498$ and we do have a bayesian effect. Also, note that $\hat{\gamma}<1$, which implies that high $Prior$'s drive an update decision, a quite reasonable result. Remember that the official homicide rate is less than 1\%, hence the higher initial responses are set, individuals are making higher mistakes, and the informational shock is strong enough to produce a change in perceptions.

The story behind $\mathbf{X}$ variables is straightforward. $Age$ has an estimated coefficient equal to $-0.728$, i.e. $\hat{\delta}_{Age} = 0.728$.  Thus, respondents' age works undoubtedly against the update: older individuals are less likely to revise initial perceptions. The same direction is found in $Gender$. Being a female drops the likelihood of a change in perceptions. Perhaps both results tell us that vulnerability faced by older citizens and women make them more reluctant to update victimization expectations.

These results are not surprisingly, though. Although fear of crime is a concern for people of all genders, studies consistently find that women around the world tend to have much higher levels of fear of crime than men (see among many, \citet{Mark1984}, \citet{Ferraro1995a}, \citet{Ferraro1996a}, \citet{Tulloch2000}, and \citet{Snedker2015}). This is so, despite the fact that for most offenses, men's actual victimization rates are higher. As to the age effect, we are in accordance to many studies that found a monotonically increasing or, at least, an ``U-shaped'' relationship between fear of crime and victimization (see, \citet{Mark1984}, and \citet{Tulloch2000}).

Despite much discussion on the importance of individuals' cognitive capability on probabilistic reasoning, our results show that education is not relevant for this task. In fact, controlling for participants' level of education brought us insignificant coefficients. We argue that above a given threshold, ensuring a basic understanding of our probability explanation, the interviewer's capacity to communicate is crucial to persuade an update. This explains a $19.307$, i.e., $\hat{\delta}_{Educ\_Int} = -19.307$  coefficient for the educational level of the informational sender. On average, graduates or undergraduates conducting our interview increase the likelihood of an update due to, we argue, a higher informational quality inherent to more educated interviewers. 

Finally, one of the most important variable in our model is $Matching\_Gender$. Controlling for individuals' gender, if the interaction is composed by the same gender, i.e. Male/Male or Female/Female, the update is more likely to occur\pagenote{Due to the \textit{ceteris paribus} assumption, we can not keep the \textit{Gender} variable in the model and identify at the same time specific matchings such as Male/Female, Female/Male and so on.}. Clearly, due to reasons out of our scope in this study, the informational content depends heavily on gender correspondence.

\subsection{Bayesian Update with Skepticism: Model 2}

As proposed in subsection (\ref{Subsection Simple Two-Tiered Hurdle Model}), we have a two-tiered model to deal with both updaters and skeptical respondents. The first stage explains the changing decision with a \textit{Logit} and a \textit{Probit} model followed by OLS on the updating subsample in the second stage. Unlike the previous subsection, the estimation results are interpreted directly. Table (\ref{Changing Decision}) and Table (\ref{Table: Restricted OLS}) present our results for the first and second stages, respectively. 

Starting for the changing decision (Table (\ref{Changing Decision})), it is important to highlight that our model is robust to the \textit{Logit} or \textit{Probit} choice. All explanatory variables are significant in the same magnitude with the same signs for both of them. Also, the Log Likelihood and the Akaike Information Criterion are almost the same. Although we can not say much on the coefficients magnitude, the signs are informative, and lead us to the conclusions made in the previous section. However, we must test our model adjustment.

So, we performed a simple exercise (results can be obtained under request) to assess the prediction power of our models, where the variable \textit{Predicted} is just our \textit{Logit} and \textit{Probit} probabilities estimation. We assume there is a threshold set at 50\% such that if $Predicted \geq 50\%$ and $Change=1$ or if $Predicted < 50\%$ and $Change=0$, then we have a correct predicted choice. Finally, the variable \textit{Correct} equals one if we had success in predicting and zero on the contrary. Therefore, \textit{Correct} mean equals $0.957$ implies that we had a $95.7\%$ success rate. 

\begin{table}[h!]
\footnotesize
\centering
\caption{Changing Decision and Marginal Effects}
\label{Changing Decision}
\begin{tabular}{@{\extracolsep{5pt}}lcccc}
\\[-1.8ex]\hline
\hline \\[-1.8ex]
& \multicolumn{2}{c}{\textit{Coefficients:}} & \multicolumn{2}{c}{\textit{Marginal effects:}} \\
\cline{2-3} \cline{4-5} 
\\[-2.0ex]
\\[-2.8ex] & Change (\textit{logistic}) & Change (\textit{probit}) & Marginal effects (\textit{logistic}) & Marginal effects (\textit{probit})\\
\hline \\[-1.8ex]
Prior & 0.007$^{**}$ & 0.003$^{**}$ &  0.0003$^{**}$ & 0.0003$^{**}$ \\
& \footnotesize{(0.003)} & \footnotesize{(0.002)} & \footnotesize{(0.0001)} & \footnotesize{(0.0001)}\\
\\[-2.5ex]
Age & $-$0.020$^{***}$ & $-$0.009$^{***}$ & $-$0.001$^{***}$ & $-$0.001$^{***}$\\
& \footnotesize{(0.007)} & \footnotesize{(0.003)} & \footnotesize{(0.0003)} & \footnotesize{(0.0002)}\\
\\[-2.5ex]
Gender & $-$0.498$^{**}$ & $-$0.238$^{**}$ & $-$0.020$^{**}$ & $-$0.021$^{**}$ \\
& \footnotesize{(0.197)} & \footnotesize{(0.094)} & \footnotesize{(0.008)} & \footnotesize{(0.008)} \\
\\[-2.5ex]
Matching\_Gender & 1.876$^{***}$ & 0.804$^{***}$ & 0.075$^{***}$ & 0.070$^{***}$\\
& \footnotesize{(0.410)} & \footnotesize{(0.162)} & \footnotesize{(0.014)} & \footnotesize{(0.013)}\\
\\[-2.5ex]
Educ\_Int & 0.654$^{*}$ & 0.261$^{*}$ & 0.026$^{**}$ & 0.023$^{*}$ \\
& \footnotesize{(0.337)} & \footnotesize{(0.141)} & \footnotesize{(0.013)} & \footnotesize{(0.012)}\\
\\[-2.5ex]
Constant & $-$5.879$^{***}$ & $-$2.813$^{***}$ & $-$0.236$^{***}$ & $-$0.244$^{***}$ \\
& \footnotesize{(1.056)} & \footnotesize{(0.436)} & \footnotesize{(0.040)} & \footnotesize{(0.037)} \\
\hline \\[-1.8ex]
\footnotesize{Observations} & \footnotesize{2,829} & \footnotesize{2,829} & & \\
\footnotesize{Log Likelihood} & \footnotesize{$-$473.495} & \footnotesize{$-$473.564}  & & \\
\footnotesize{Akaike Inf. Crit.} & \footnotesize{958.990} & \footnotesize{959.129}  & & \\
\hline
\hline \\[-1.8ex]
\multicolumn{3}{l}{\footnotesize{$^{*}$p$<$0.1; $^{**}$p$<$0.05; $^{***}$p$<$0.01}}\\
\multicolumn{3}{l}{\footnotesize{Source: Elaborated by the author.}}\\
\end{tabular}
\end{table}

Given our good adjustment, we proceed to interpret its coefficients. Table (\ref{Changing Decision}) presents the marginal effects of our explanatory variables over the changing decision probability. With respect to the \textit{Prior}, in accordance with our previous model, initial perceptions have a positive influence on the probability that an update occurs. Higher initial perceptions imply higher mistakes, hence, once more, this willingness to change can be taken as a rational decision.

Also, for all of the other independent variables, the results lead to the same previous conclusions. In \textit{Age}, as it was the case before, older participants are more reluctant to change and, again, perhaps insecurity drives this results. A stronger evidence of this explanation comes from \textit{Gender}. Being female reduces the changing probability in $2$ percentage points.  

As expected, educated informational senders had a positive impact on the changing decision. Here, clearness or even social aspects of education in a low-educated society might influence the information quality and credibility. Once again, homophily analysis arises. Being of the same gender as the interviewer is clearly the most important aspect in our data. Male respondents might consider reliable only information provided by male interviewers. On the other hand, female participants might feel more comfortable being interviewed by women. Many explanations are possible but, under this cultural context, it is intuitive that gender matchings reduce information noise and this is key to the informational content. 

\begin{table}[h!]
\footnotesize
\centering
	\caption{Restricted Ols}
	\label{Table: Restricted OLS}
	\begin{tabular}{@{\extracolsep{5pt}}lc}
		\\[-1.8ex]\hline
		\hline \\[-1.8ex]
		& \multicolumn{1}{c}{\textit{Dependent variable:}} \\
		\cline{2-2}
		\\[-1.8ex] & Post \\
		\hline \\[-1.8ex]
		Prior & 0.023$^{***}$ \\
		& \footnotesize{(0.008)} \\
		   \\[-2.5ex]
		Age & 0.026$^{*}$ \\
		& \footnotesize{(0.015)} \\
		   \\[-2.5ex]
		Gender & 0.988$^{**}$ \\
		& \footnotesize{(0.455)} \\
		   \\[-2.5ex]
		Matching\_Gender & 1.074 \\
		& \footnotesize{(1.179)} \\
		   \\[-2.5ex]
		Educ\_Int & $-$4.188$^{***}$ \\
		& \footnotesize{(0.975)} \\
		  \\[-2.5ex]
		Police & 0.494$^{*}$ \\
		& \footnotesize{(0.273)} \\
		   \\[-2.5ex]
		Constant & 1.697 \\
		& \footnotesize{(1.369)} \\
		\hline \\[-1.8ex]
		\footnotesize{Observations} & \footnotesize{121} \\
		\footnotesize{Adjusted R$^{2}$} & \footnotesize{0.254} \\
		\footnotesize{Residual Std. Error} & \footnotesize{2.375 (df = 114)}\\
		\footnotesize{F Statistic} & \footnotesize{7.822$^{***}$ (df = 6; 114)}\\
		\hline
		\hline \\[-1.8ex]
		\multicolumn{2}{l}{\footnotesize{$^{*}$p$<$0.1; $^{**}$p$<$0.05; $^{***}$p$<$0.01}}\\
		\multicolumn{2}{l}{\footnotesize{Source: Elaborated by the author.}}\\
	\end{tabular}
\end{table}

Once having initial perceptions changed, now we proceed to the updating analysis, see Table \ref{Table: Restricted OLS}. We have two key points to emphasize with this exercise: in line with our Tobit model and with \citet{Viscusi1984}, the \textit{Prior} coefficient in a linear model is significant. This is a crucial evidence supporting our Bayesian approach both for the general and restricting cases. \textit{Age} and \textit{Gender} have exactly the same role as our previous analysis and there is no need to say anything else. In the same way, the interviewers' level of education has again an important role explaining posterior responses. As expected, since all \textit{Post}'s are less than \textit{Prior}'s, and these initial perceptions are more than 1000 times greater than the true value, more educated interviewers move participants closer to the truth. This explains a negative value for \textit{Educ\_Int} ($-4.188$).

The insignificance of $Matching\_Gender$ is new here. However, remember from table (\ref{Descriptive Analysis - Updating Subsample}) that $93\%$ of our updating subsample is composed by interactions of the same gender. Hence, perhaps we do not have variations enough to obtain a significant coefficient which means that this pre-evidence is just what we are looking for. 

Also, given the flexibility of the two-tiered modeling, we could introduce another independent variable: \textit{Police}. Its positive coefficient is just what one would expect. Higher levels of \textit{Police} values indeed mean a worse evaluation. Therefore, it leads us to an intuitive result: posterior perceptions are higher or, in other words, the movement towards the truth is less strong, when police work is worse seen by respondents. 

\section{FINAL CONSIDERATIONS} \label{Section: CONCLUSIONS}	

This paper was built under three major guidances: i) Individuals are rational decision makers when updating subjective perceptions; ii) There is an estimable linear regression model for the bayesian update process, a well-suited framework to deal with the revision of subjective perceptions; iii) There are other variables besides respondents' characteristics influencing this updating process. 

With these ideas, we presented an initial sample of 4,030 individuals regarding subjective risk perceptions about becoming a homicide victim for the following 12 months in Fortaleza, Brazil. Besides having a much larger sample than those presented in many papers in the field, our data brought information about interviewers and matching aspects that were used to account for information quality. In addition, we had a very different data sample: $95\%$ of respondents did not want to change their \textit{Prior's}, setting $Prior_{i} = Post_{i}$. This made the non-updaters, or skeptical agents, our protagonists. 

We showed that, under simple assumptions, \citet{Viscusi1984} and \citet{Smith1988} could be extended at the same time. We proposed a multiple linear regression model in the context of a bayesian update approach using a vector of independent variables for respondents, interviewers and matching in gender. However, our empirical evidence imposed the development of a model able to accommodate both updaters and non-updaters. 

Following this guidance, we proposed two alternative estimable models suited to a more general context than what is found in the literature. Indeed, we believe this generalization to be a simple, yet important, achievement of this study. A modified Tobit, easily implemented in any statistical package, was developed as the first approach. The second model was a well-known two-tiered Hurdle model, allowing the possibility to use a different set of variables to explain the changing decision and the update.

Firstly, our results showed that we could proceed with a bayesian update approach, since our $Prior$ coefficient was significant for every model used. More specifically, referring to our Tobit model, the initial response coefficient was $\hat{\gamma} = (1-0.5021) = 0.498$, and it was statistically significant. It permitted us to conclude that $Prior's$ induce an update through a latent variable. In other words, higher initial perceptions imply higher misleading perceptions and this leads to a change in responses. Despite the fact that we are using a different framework, our results stay in accordance with that of \citet{Viscusi1984}, i.e. there is a statistically significant linear equation relating \textit{priors} to \textit{posteriors}.

Also, fundamentally, we could rationalize a non-updating decision following a perceived informational quality/credibility argument. This was made through respondents' age, gender and initial perceptions, as well as interviewers' level of education and matching in gender between members of the interaction. 

We found that older participants and females are more reluctant not only to change initial responses, but also to choose the level of the new response, in case of an update. We argued that insecurity aspects might be used to explain these findings. Also, respondents' level of education was insignificant in our exercise. In fact, interviewers' level of education plays a key role in both the changing and updating magnitude decisions. This is a consistent piece of evidence supporting our main line of explanation: the perceived informational content. The relation between sender's education and information quality is straightforward. 

Finally, our results also raised important evidence on homophily aspects. In almost every regression, $Matching\_Gender$ had a major impact on the decision to change and in the magnitude of the update. The only insignificant one was in a simple OLS on the updating subsample. However, in this group, $93\%$ of the interactions happened under the same gender. It suggests that Latin cultures, as the Brazilian, do not pose weight on race questions, as it is the case for other cultures. However, it has an analogue restriction due, perhaps, to its characteristic machismo: gender. 

Further developments should be done in many aspects of this study. Firstly, the design of our interaction was clearly not ideal and the informational shock could be better customized to adjust official rates to different socioeconomic characteristics. Although it raised a remarkable opportunity to propose a generalization, our estimation procedure can be improved in several ways. One of particular importance is to allow different revision directions, since our data and modeling present just the downwards update. Also, we could have more information regarding interviewers. Further study on homophily is another gap to be filled in. 


\bibliographystyle{apalike}
\bibliography{Dissertation}

\begin{thebibliography}{}

\bibitem[Becker, 1968]{Becker1968}
Becker, C. (1968).
\newblock Punishment: An economic approach, 76j.
\newblock {\em Pol. Econ}, 169(10.2307):1830482169.

\bibitem[Bott, 1928]{Bott1928}
Bott, H. (1928).
\newblock Observation of play activities in a nursery school.
\newblock {\em Genetic Psychology Monographs}, 4(1):44--88.

\bibitem[Bursik~Jr et~al., 1999]{BursikJr1999}
Bursik~Jr, R.~J., Grasmick, H.~G., et~al. (1999).
\newblock {\em Neighborhoods \& crime}.
\newblock Lexington Books.

\bibitem[Delavande, 2008]{Delavande2008}
Delavande, A. (2008).
\newblock Measuring revisions to subjective expectations.
\newblock {\em Journal of Risk and Uncertainty}, 36(1):43--82.

\bibitem[Delavande et~al., 2011]{Delavande2011}
Delavande, A., Gin{\'e}, X., and McKenzie, D. (2011).
\newblock Measuring subjective expectations in developing countries: A critical
  review and new evidence.
\newblock {\em Journal of Development Economics}, 94(2):151--163.

\bibitem[Dominitz and Manski, 1996]{Dominitz1996}
Dominitz, J. and Manski, C.~F. (1996).
\newblock Perceptions of economic insecurity: Evidence from the survey of
  economic expectations.
\newblock Technical report, National bureau of economic research.

\bibitem[Dominitz and Manski, 1997]{Dominitz1997}
Dominitz, J. and Manski, C.~F. (1997).
\newblock Using expectations data to study subjective income expectations.
\newblock {\em Journal of the American Statistical Association},
  92(439):855--867.

\bibitem[Dominitz and Manski, 2004]{Dominitz2004}
Dominitz, J. and Manski, C.~F. (2004).
\newblock How should we measure consumer confidence?
\newblock {\em Journal of Economic Perspectives}, pages 51--66.

\bibitem[Dominitz and Manski, 2011]{Dominitz2011}
Dominitz, J. and Manski, C.~F. (2011).
\newblock Measuring and interpreting expectations of equity returns.
\newblock {\em Journal of Applied Econometrics}, 26(3):352--370.

\bibitem[Dominitz et~al., 2002]{Dominitz2002}
Dominitz, J., Manski, C.~F., and Heinz, J. (2002).
\newblock Social security expectations and retirement savings decisions.
\newblock Technical report, National Bureau of Economic Research.

\bibitem[El-Gamal and Grether, 1995]{El-Gamal1995}
El-Gamal, M.~A. and Grether, D.~M. (1995).
\newblock Are people bayesian? uncovering behavioral strategies.
\newblock {\em Journal of the American statistical Association},
  90(432):1137--1145.

\bibitem[Ferraro, 1995a]{Ferraro1995}
Ferraro, K.~F. (1995a).
\newblock {\em Fear of Crime: Interpreting Victimization Risk}.
\newblock State University of New Yoor Press.

\bibitem[Ferraro, 1995b]{Ferraro1995a}
Ferraro, K.~F. (1995b).
\newblock {\em Fear of crime: Interpreting victimization risk}.
\newblock SUNY press.

\bibitem[Ferraro, 1996]{Ferraro1996a}
Ferraro, K.~F. (1996).
\newblock Women's fear of victimization: Shadow of sexual assault?
\newblock {\em Social forces}, pages 667--690.

\bibitem[Ferrell and McGoey, 1980]{Ferrell1980}
Ferrell, W.~R. and McGoey, P.~J. (1980).
\newblock A model of calibration for subjective probabilities.
\newblock {\em Organizational Behavior and Human Performance}, 26(1):32--53.

\bibitem[Golub and Jackson, 2012]{Golub2012}
Golub, B. and Jackson, M.~O. (2012).
\newblock Network structure and the speed of learning measuring homophily based
  on its consequences.
\newblock {\em Annals of Economics and Statistics/ANNALES D'{\'E}CONOMIE ET DE
  STATISTIQUE}, pages 33--48.

\bibitem[Graham, 2011]{Graham2011}
Graham, B. (2011).
\newblock Econometric methods for the analysis of assignment problems in the
  presence of complementarity and social spillovers.
\newblock {\em Handbook of social economics}, 1:965--1052.

\bibitem[Hart et~al., 1960]{Hart1960}
Hart, A.~G., Modigliani, F., and Orcutt, G.~H. (1960).
\newblock The quality and economic significance of anticipations data.

\bibitem[Hoffrage et~al., 2000]{Hoffrage2000}
Hoffrage, U., Lindsey, S., Hertwig, R., and Gigerenzer, G. (2000).
\newblock Communicating statistical information.
\newblock {\em Science}, 290(5500):2261--2262.

\bibitem[Hurd and McGarry, 1995]{Hurd1995}
Hurd, M.~D. and McGarry, K. (1995).
\newblock Evaluation of the subjective probabilities of survival in the health
  and retirement study.
\newblock {\em Journal of Human resources}, pages S268--S292.

\bibitem[Jackson, 2009]{Jackson2009}
Jackson, M.~O. (2009).
\newblock Social structure, segregation, and economic behavior.
\newblock {\em Segregation, and Economic Behavior (February 5, 2009)}.

\bibitem[Juster, 1966]{Juster1966}
Juster, F.~T. (1966).
\newblock Consumer buying intentions and purchase probability: An experiment in
  survey design.
\newblock {\em Journal of the American Statistical Association},
  61(315):658--696.

\bibitem[Koriat et~al., 1980]{Koriat1980}
Koriat, A., Lichtenstein, S., and Fischhoff, B. (1980).
\newblock Reasons for confidence.
\newblock {\em Journal of Experimental Psychology: Human learning and memory},
  6(2):107.

\bibitem[Kunreuther et~al., 1978]{Kunreuther1978}
Kunreuther, H., Ginsberg, R., Miller, L., Sagi, P., Slovic, P., Borkan, B., and
  Katz, N. (1978).
\newblock {\em Disaster insurance protection: Public policy lessons}.
\newblock Wiley New York.

\bibitem[Lichtenstein et~al., 1978]{Lichtenstein1978}
Lichtenstein, S., Slovic, P., Fischhoff, B., Layman, M., and Combs, B. (1978).
\newblock Judged frequency of lethal events.
\newblock {\em Journal of experimental psychology: Human learning and memory},
  4(6):551.

\bibitem[Machlup, 1946]{Machlup1946}
Machlup, F. (1946).
\newblock Marginal analysis and empirical research.
\newblock {\em The American Economic Review}, pages 519--554.

\bibitem[Manski, 2004]{Manski2004}
Manski, C.~F. (2004).
\newblock Measuring expectations.
\newblock {\em Econometrica}, 72(5):1329--1376.

\bibitem[Mark, 1984]{Mark1984}
Mark, W. (1984).
\newblock Fear of victimization: Why are women and the elderly more afraid?
\newblock {\em Social science quarterly}, 65(3):681.

\bibitem[McFadden, 1973]{McFadden1973}
McFadden, D. (1973).
\newblock Conditional logit analysis of qualitative choice behavior.

\bibitem[McPherson et~al., 2001]{McPherson2001}
McPherson, M., Smith-Lovin, L., and Cook, J.~M. (2001).
\newblock Birds of a feather: Homophily in social networks.
\newblock {\em Annual review of sociology}, pages 415--444.

\bibitem[Pesaran, 1987]{Pesaran1987}
Pesaran, M.~H. (1987).
\newblock {\em The limits to rational expectations}.
\newblock Blackwell Oxford.

\bibitem[Samuelson, 1938]{Samuelson1938}
Samuelson, P.~A. (1938).
\newblock A note on the pure theory of consumer's behaviour.
\newblock {\em Economica}, pages 61--71.

\bibitem[Smith and Johnson, 1988]{Smith1988}
Smith, V.~K. and Johnson, F.~R. (1988).
\newblock How do risk perceptions respond to information? the case of radon.
\newblock {\em The Review of Economics and Statistics}, 70(1):1--8.

\bibitem[Snedker, 2015]{Snedker2015}
Snedker, K.~A. (2015).
\newblock Neighborhood conditions and fear of crime a reconsideration of sex
  differences.
\newblock {\em Crime \& Delinquency}, 61(1):45--70.

\bibitem[Tulloch, 2000]{Tulloch2000}
Tulloch, M. (2000).
\newblock The meaning of age differences in the fear of crime: Combining
  quantitative and qualitative approaches.
\newblock {\em The British Journal of Criminology}, pages 451--467.

\bibitem[Tversky and Kahneman, 1974]{Tversky1974}
Tversky, A. and Kahneman, D. (1974).
\newblock Judgment under uncertainty: Heuristics and biases.
\newblock {\em science}, 185(4157):1124--1131.

\bibitem[{United Nations}, 2013]{Drugs2013}
{United Nations} (2013).
\newblock {\em Global Study on Homicide 2013: Trends, Contexts, Data}.

\bibitem[Vanderveen, 2006]{Vanderveen2006}
Vanderveen, G. (2006).
\newblock {\em Interpreting Fear, Crime, Risk and Unsafety: Conceptualisation
  and Measurement}.
\newblock Boom Juridische Uitgevers.

\bibitem[Viscusi, 1979]{Viscusi1979}
Viscusi, W.~K. (1979).
\newblock {\em Employment hazards: An investigation of market performance}.
\newblock Number 148. Harvard University Press.

\bibitem[Viscusi, 1985]{Viscusi1985}
Viscusi, W.~K. (1985).
\newblock Are individuals bayesian decision makers?
\newblock {\em The American Economic Review}, pages 381--385.

\bibitem[Viscusi and O'Connor, 1984]{Viscusi1984}
Viscusi, W.~K. and O'Connor, C.~J. (1984).
\newblock Adaptive responses to chemical labeling: Are workers bayesian
  decision makers?
\newblock {\em American Economic Review}, 74(5):942--956.

\bibitem[Warr, 2000]{Warr2000}
Warr, M. (2000).
\newblock Fear of crime in the united states: Avenues for research and policy.
\newblock Technical report, National Institute of Justice/NCJRS, Rockville, MD
  - USA.

\end{thebibliography}

\newpage
\renewcommand*{\notedivision}{\section*{\notesname}}

\printnotes

\end{document}